\documentclass[final,copyright,creativecommons]{eptcs}
\providecommand{\event}{GandALF 2016}
\hypersetup{colorlinks=true,citecolor=blue,linkcolor=red}
\usepackage[utf8]{inputenc}
\usepackage{hhline}

\usepackage{xcolor,tikz}
\usetikzlibrary{arrows,automata,shapes,backgrounds,decorations.pathreplacing}

\usepackage{amsmath,amsthm}
\usepackage{amssymb,stmaryrd,mathtools}
\usepackage{enumitem}
\usepackage{multirow}

\newtheorem{lemma}{Lemma}
  \newtheorem{definition}{Definition}
  \newtheorem{theorem}{Theorem}
  
  \newtheorem{proposition}{Proposition}
  \theoremstyle{remark}
  
  \newtheorem{example}{Example}
\makeatletter
\let\c@definition\c@theorem
\let\c@lemma\c@theorem
\let\c@corollary\c@theorem
\let\c@remark\c@theorem
\let\c@example\c@theorem
\let\c@proposition\c@theorem
\makeatother

\title{Window Parity Games: An Alternative Approach Toward Parity Games with Time Bounds\thanks{Q.~Hautem is supported by a FRIA fellowship, M.~Randour is an F.R.S.-FNRS Postdoctoral researcher.}}

\author{V\'eronique Bruy\`ere \qquad Quentin Hautem\institute{Computer Science Department, Université de Mons (UMONS), Belgium}\\Mickael Randour\institute{Computer Science Department, Universit\'e libre de Bruxelles (ULB), Belgium\\Computer Science Department, Université de Mons (UMONS), Belgium}}	 

\def\titlerunning{Window Parity Games}
\def\authorrunning{V.~Bruy\`ere, Q.~Hautem \& M.~Randour}

\newcommand{\N}{\mathbb{N}}

\newcommand{\Nzero}{\N_0}

\newcommand{\ssetminus}{\! \setminus \!}

\newcommand{\stronger}{\preceq}

\newcommand{\smallest}{$\stronger$-smallest}
\newcommand{\smaller}{$\stronger$-smaller}

\newcommand{\playerOne}{\ensuremath{\mathcal{P}_1} } 
\newcommand{\playerTwo}{\ensuremath{\mathcal{P}_2 } }
\newcommand{\playerI}{\ensuremath{\mathcal{P}_i} }
\newcommand{\Gdev}{(V_1,V_2, E)}   
\newcommand{\Plays}{\mathsf{Plays}}

\newcommand{\Out}{\mathsf{Out}}
\newcommand{\Obj}{\Omega}  
\newcommand{\WinG}[3]{{\mathsf{Win}}_{#1}^{#3}({#2})}  

\newcommand{\pifactor}[2]{\pi{[{#1},{#2}]}}

\newcommand{\Reach}{\mathsf{Reach}}
\newcommand{\Safe}{\mathsf{Safe}}
\newcommand{\Buchi}{\mathsf{Buchi}}
\newcommand{\CoBuchi}{\mathsf{CoBuchi}}
\newcommand{\GenReach}{\mathsf{GenReach}}

\newcommand{\Par}{\mathsf{Parity}}

\newcommand{\Inf}{\mathsf{Inf}}

\newcommand{\DirFWP}{\mathsf{DirFixWP}}
\newcommand{\FWP}{\mathsf{FixWP}}
\newcommand{\DirBWP}{\mathsf{DirBndWP}}
\newcommand{\BWP}{\mathsf{BndWP}}
\newcommand{\DirBP}{\mathsf{DirBndPR}}
\newcommand{\BP}{\mathsf{BndPR}}
\newcommand{\DirPR}{\mathsf{DirFixPR}}
\newcommand{\PR}{\mathsf{FixPR}}
\newcommand{\RR}{\mathsf{RR}}

\newcommand{\BndRR}{\mathsf{BndRR}}
\newcommand{\DirBndRR}{\mathsf{DirBndRR}}

\newcommand{\window}[1]{window at position~${#1}$}

\newcommand{\ffclosed}[1]{first-closed}

\begin{document}
  \maketitle	

\begin{abstract} 
Classical objectives in two-player zero-sum games played on graphs often deal with limit behaviors of infinite plays: e.g., \textit{mean-payoff} and \textit{total-payoff} in the quantitative setting, or \textit{parity} in the qualitative one (a canonical way to encode $\omega$-regular properties). Those objectives offer powerful abstraction mechanisms and often yield nice properties such as memoryless determinacy.
However, their very nature provides no guarantee on time bounds within which something good can be witnessed. 
In this work, we consider two approaches toward inclusion of \textit{time bounds} in parity games. The first one, \textit{parity-response} games, is based on the notion of finitary parity games~\cite{ChatterjeeHH09} and parity games with costs~\cite{DBLP:journals/corr/abs-1207-0663,Weinert016}. The second one, \textit{window parity} games, is inspired by window mean-payoff games~\cite{Chatterjee0RR15}. We compare the two approaches and show that while they prove to be equivalent in some contexts, window parity games offer a more tractable alternative when the time bound is given as a parameter ($\sf P$-c.~vs. $\sf PSPACE$-c.). In particular, it provides a conservative approximation of parity games computable in polynomial time. Furthermore, we extend both approaches to the multi-dimension setting. We give the full picture for both types of games with regard to complexity and memory bounds.
\end{abstract}

\section{Introduction}
\label{sec:intro}

\smallskip\noindent\textbf{Games on graphs.} Two-player games played on directed graphs constitute an important framework for the synthesis of a suitable controller for a reactive system faced to an uncontrollable environment~\cite{randourECCS}. In this setting, \textit{vertices} of the graph represent states of the system and edges represent transitions between those states. We consider \textit{turn-based two-player} games: each vertex either belongs to the system (the first player, denoted by~$\playerOne$) or the environment (the second player, denoted by $\playerTwo$). A game is played by moving an imaginary pebble from vertex to vertex according to existing transitions: the owner of a vertex decides where to move the pebble. The outcome of the game is an infinite sequence of vertices called \textit{play}. The choices of both players depend on their respective \textit{strategy} which can use an arbitrary amount of memory in full generality. In the classical setting, $\playerOne$ tries to achieve an \textit{objective} (describing a set of \textit{winning plays}) while $\playerTwo$ tries to prevent him from succeeding: hence, our games are \textit{zero-sum}. As all the objectives considered in this paper define Borel sets, Martin's theorem~\cite{Martin75} guarantees determinacy.

\smallskip\noindent\textbf{Parity games.} Two-player games with $\omega$-regular objectives have been studied extensively in the literature. See for example~\cite{Thomas97,2001automata} for an introduction. A canonical way to represent games with $\omega$-regular conditions is the class of \textit{parity games}: vertices are assigned a non-negative integer \textit{priority} (or \textit{color}), and the objective asks that among the vertices that are seen infinitely often along a play, the minimal priority be even. Parity games have been under close scrutiny for a long time both due to their importance (e.g., they subsume modal $\mu$-calculus model checking~\cite{DBLP:conf/cav/EmersonJS93}) and their intriguing complexity: they belong to the class of problems in $\sf UP \cap coUP$~\cite{Jurdzinski98} and despite many efforts (e.g.,~\cite{Zielonka98,Jurdzinski00,DBLP:journals/siamcomp/JurdzinskiPZ08,DBLP:conf/fsttcs/Schewe07}), whether they belong to $\sf P$ is still an open question. Furthermore, parity games enjoy memoryless determinacy~\cite{EmersonJ88,Zielonka98}. Multi-dimension parity games were studied in~\cite{DBLP:conf/fossacs/ChatterjeeHP07}: in such games, $n$-dimension vectors of priorities are associated to each vertex, and the objective is to satisfy the conjunction of all the one-dimension parity objectives. The complexity of solving those games is higher: deciding if $\playerOne$ (resp.~$\playerTwo$) has a winning strategy is $\sf coNP$-complete (resp.~$\sf NP$-complete) and exponential memory is needed for $\playerOne$ whereas $\playerTwo$ remains memoryless~\cite{DziembowskiJW97,Hor05-GDV}.

\smallskip\noindent\textbf{Time bounds.} In its classical formulation, the parity objective essentially requires that \textit{for each odd priority seen infinitely often, a smaller even priority should also be seen infinitely often}. An odd priority can be seen as a stimulus that must be answered by seeing a smaller even priority. The parity objective has fundamental qualities. The simplicity of its definition abstracts timing issues like ``how much time has elapsed between a stimulus and its answer'' and is key to memoryless determinacy. It makes it robust to slight changes in the model which could impact more precise formulations (e.g., counting the number of steps between a stimulus and its answer critically depends on the granularity of the game graph).

Nonetheless, it has been recently argued that in a large number of practical applications, \textit{timing does matter} (e.g.,~\cite{ChatterjeeHH09,DBLP:journals/fmsd/KupfermanPV09,Chatterjee0RR15}). Indeed, in general it does not suffice to know that a ``good behavior'' will \textit{eventually} happen, and one wants to ensure that it can actually be witnessed \textit{within a time frame which is acceptable} with regard to the modeled reactive system. For example, consider a computer server having to grant requests to clients. A classical parity objective can encode that requests should eventually be granted. However, it is clear that in a desired controller, requests should not be placed on hold for an arbitrarily long time. In order to accomodate such requirements, various attempts to associate classical game objectives with time bounds have been recently studied. For example, \textit{window mean-payoff} and \textit{window total-payoff} games provide a framework to reason about quantitative games (e.g., modeling quantities such as energy consumption) with time bounds~\cite{Chatterjee0RR15}. In the qualitative setting, \textit{finitary parity} games~\cite{ChatterjeeHH09,ChatterjeeF13} and \textit{parity games with costs}~\cite{DBLP:journals/corr/abs-1207-0663,Weinert016} provide a similar framework for parity games.

\smallskip\noindent\textbf{Two approaches.} While window games and finitary parity games (resp.~parity games with costs) share the goal of allowing precise specification of time bounds, their inner mechanisms differ. The aim of our work is three-fold: $(i)$~apply the \textbf{window mechanism to parity games}, $(ii)$~provide a \textbf{thorough comparison} with the existing framework of finitary parity games and parity games with costs, $(iii)$~\textbf{extend both approaches to the multi-dimension setting} (which was left unexplored up to now). Since all those related papers do not use a uniform terminology, we here use the following taxonomy for the two approaches.
\begin{itemize}
\item \textbf{Window parity (WP).} Intuitively, the \textit{direct fixed WP} objective considers a window of size bounded by $\lambda \in \mathbb{N}_0$ (given as a parameter) sliding over an infinite play and declare this play winning if in all positions, the window is such that the minimal priority within it is even. For \textit{direct bounded WP}, the size of the window is not fixed as a parameter but a play is winning if there exists a bound $\lambda$ for which the condition holds. We also consider the \textit{fixed WP} and \textit{bounded WP} objectives which are essentially prefix-independent variants of the previous ones. All those objectives are based on the window mechanism introduced in~\cite{Chatterjee0RR15} and our work presents the first implementation of this mechanism for parity games.

\item \textbf{Parity-response (PR).} The \textit{direct fixed PR} objective asks that along a play, any odd priority be followed by a smaller even priority in at most $\lambda \in \mathbb{N}_0$ (given as a parameter) steps. As for the WP setting, we also consider the \textit{direct bounded PR} objective where a play is winning if there exists a bound~$\lambda$ such that the condition holds, along with the respective prefix-independent variants: the \textit{fixed PR} and the \textit{bounded PR} objectives. The \textit{bounded PR} objective was studied for one-dimension games (under the name \textit{finitary parity}) in~\cite{ChatterjeeHH09}: deciding the winner is in $\sf P$ and memoryless strategies suffice for $\playerOne$ while $\playerTwo$ needs infinite memory. The \textit{fixed PR} objective for one-dimension games was very recently proved to be $\sf PSPACE$-complete, with exponential memory bounds for both players~\cite{Weinert016} (this work is presented in the more general context of \textit{parity games with costs}). Our work provides the first study of the parity-response approach in multi-dimension games.
\end{itemize}

\renewcommand{\arraystretch}{1.4}
\begin{table}
\begin{center}
\scalebox{0.8}{\begin{tabular}{|c||c|c|c||c|c|c|}
\cline{2-7}
\multicolumn{1}{c|}{} & \multicolumn{3}{c||}{one-dimension} & \multicolumn{3}{c|}{ multi-dimension} \\
\cline{2-7}
\multicolumn{1}{c|}{} & ~complexity~ & $\playerOne$ mem. & $\playerTwo$ mem. & ~complexity~ & $\playerOne$ mem. & $~\playerTwo$ mem.$~$ \\
\hhline{-|======|}
\textbf{Fixed WP} & \textbf{\textsf{P}-c.} & \multicolumn{2}{c||}{\textbf{polynomial}}  & ~\multirow{4}{*}{\textbf{\textsf{EXPTIME}-c.}}~ & \multicolumn{2}{c|}{\multirow{2}{*}{\textbf{exponential}}} \\
\cline{1-4}
Fixed PR & \textsf{PSPACE}-c. & ~exponential~ & ~$\leq$ exponential~ & & \multicolumn{2}{c|}{} \\
\cline{1-4} \cline{6-7}
~\textbf{Bounded WP}~ & \multirow{2}{*}{\textsf{P}-c.} & \multirow{2}{*}{memoryless} & \multirow{2}{*}{infinite} & & \multirow{2}{*}{$~$\textbf{exponential}$~$} & \multirow{2}{*}{\textbf{infinite}} \\ 
\cline{1-1}
~Bounded PR~ & & & & & & \\
\cline{1-7}
\end{tabular}}
\end{center}
\vspace{-3mm}
\caption{Complexity of deciding the winner and memory required for winning strategies in window parity (WP) and parity-response (PR) games. All results hold for both the \textit{prefix-independent} and the \textit{direct} ({\sf Dir}) variants of all the objectives, except for the memory of $\playerTwo$ in the direct bounded cases: in one-dimension games, linear memory is both sufficient and necessary (for both WP and PR) and in multi-dimension games, exponential memory is both sufficient and necessary. All bounds are tight unless specified by the $\leq$ symbol. New results are in bold.}
\label{table:overview}
\end{table}

\smallskip\noindent\textbf{Our contributions.} Given the number of variants studied, we give an overview of our results in Table~\ref{table:overview}. Our main contributions are as follows. 
\begin{enumerate}
\item We prove that \textit{bounded WP} and \textit{bounded PR} objectives coincide, even in multi-dimension games (Proposition~\ref{prop:inclusions}).
\item We establish that \textit{bounded WP} (and thus \textit{bounded PR}) games are $\sf P$-hard in one-dimension (Theorem~\ref{thm:finitaryparity}, $\sf P$-membership follows from~\cite{ChatterjeeHH09}) and that they are $\sf EXPTIME$-complete in multi-dimension (Theorem~\ref{prop:multiDirBP}). The $\sf EXPTIME$-membership follows from a reduction to a variant of \textit{request-response games}~\cite{WallmeierHT03} presented in~\cite{ChatterjeeHH09} under the name of \textit{finitary Streett games}. The $\sf EXPTIME$-hardness is proved via a reduction from the membership problem in alternating polynomial-space Turing machines.
\item We show that in multi-dimension \textit{bounded WP} (and thus \textit{bounded PR}) games, exponential memory is both sufficient and necessary for $\playerOne$ while infinite memory is needed for $\playerTwo$ (Theorem~\ref{prop:multiDirBP}). 
\item We prove that one-dimension \textit{fixed WP} games provide a \textbf{conservative approximation of parity games} (Proposition~\ref{prop:inclusions}) \textbf{computable in polynomial time} (Theorem~\ref{prop:WPonedim}). This is in contrast to the $\sf PSPACE$-completeness of \textit{fixed PR} games~\cite{Weinert016} (actually, the proof in~\cite{Weinert016} is for a more general model but already holds for \textit{fixed PR} games). 
\item While \textit{fixed PR} games are $\sf PSPACE$-complete, we establish two polynomial-time algorithms (Theorem~\ref{thm:fixed}) to solve fixed-parameter sub-cases: $(i)$ the bound $\lambda$ is fixed, or $(ii)$ the number of priorities is fixed.
\item In multi-dimension, we prove that both \textit{fixed PR} (Theorem~\ref{prop:multiPR}) and \textit{fixed WP} (Theorem~\ref{prop:multiWP}) games are $\sf EXPTIME$-complete. Membership relies on different techniques and algorithms for each case while hardness is based on the same reduction as for the \textit{bounded} variants.
\item In one-dimension games, we also establish that for \textit{fixed WP}, polynomial memory is both sufficient and necessary for both players, whereas exponential memory is required for \textit{fixed PR}~\cite{Weinert016}. In multi-dimension games, we prove that for both \textit{fixed PR} and \textit{fixed WP}, exponential memory is both sufficient and necessary for both players. The upper bounds follow from the $\sf EXPTIME$ algorithms mentioned above whereas the lower bounds in one-dimension are shown thanks to appropriate families of games and in multi-dimension are obtained through reduction from \textit{generalized reachability games}~\cite{FijalkowH13}.
\item We establish the existence of values of $\lambda$ such that the \textit{fixed} objectives become equivalent to the \textit{bounded} ones, both in one-dimension (Theorem~\ref{thm:finitaryparity}) and in multi-dimension (Theorem~\ref{prop:multiDirBP}).
\end{enumerate}
While all the aforementioned results are for the \textit{prefix-independent} variants of our objectives, we also obtain closely related complexities and memory bounds for the \textit{direct} ones (Table~\ref{table:overview}). We obtain our results using a variety of techniques, sometimes inspired by~\cite{ChatterjeeHH09,Chatterjee0RR15}. Our focus is on giving the full picture for the two approaches toward including time bounds in parity games: window parity and parity-response. We sum up the key comparison points in the next paragraph.

\smallskip\noindent\textbf{Comparison.} The \textit{parity-response} and \textit{window parity} approaches turn out to be equivalent in the \textit{bounded} context, i.e., when the question is the existence of a bound $\lambda \in \mathbb{N}_0$ for which the corresponding \textit{fixed} objective holds. Hence, the focus of the comparison is the \textit{fixed} variants. Observe that those variants are of interest for applications where the time bound is part of the specification: parameter $\lambda$ grants flexibility in the specification as it can be adjusted to specific requirements of the application. Let us review the complexities of the \textit{fixed PR} and \textit{fixed WP} objectives.

In one-dimension games, \textit{fixed PR} is $\sf PSPACE$-complete whereas \textit{\textit{fixed WP} provides a framework with similar flavor that enjoys increased tractability}: it is $\sf P$-complete. Hence, \textit{\textit{fixed WP} does provide a polynomial-time conservative approximation of parity games} (Proposition~\ref{prop:inclusions}). Interestingly, the \textit{fixed WP} objective also permits to approximate the \textit{fixed PR} one in both directions, and in polynomial time: we prove in Proposition~\ref{prop:inclusions} that the \textit{fixed PR} objective for time bound $\lambda$ can be framed by the \textit{fixed WP} objective for two well-chosen values of the time bound $\lambda'$ and $\lambda''$.

In multi-dimension, both \textit{fixed PR} and \textit{fixed WP} games are $\sf EXPTIME$-complete. Still, while the \textit{fixed PR} algorithm requires exponential time in \textit{both} the number of dimensions \textit{and} the number of priorities (which can be as large as the game graph), \textit{solving the \textit{fixed WP} case only requires exponential time in the number of dimensions}. This distinction may have \textit{impact on practical applications} where, usually, the size of the model (the game graph) can be very large while the specification (hence the number of dimensions) is comparatively small. Note that for both objectives, the multi-dimension algorithms are \textit{pseudo-}polynomial in the time bound $\lambda$, hence also exponential in the length of its binary encoding.

Finally, let us compare \textit{window parity} games with \textit{window mean-payoff (WMP)} games~\cite{Chatterjee0RR15}. First, one could naturally wonder if WP games could be solved by encoding them into WMP games, following a reduction similar in spirit to the one developed by Jurdzinski for classical parity games~\cite{Jurdzinski98}. This is indeed possible, but leads to increased complexities in comparison to the ad hoc analysis developed in this work. For example, multi-dimension \textit{fixed WP} games would require exponential time in the number of priorities too. Second, observe that \textit{fixed WP} games can be solved in polynomial time whatever the bound $\lambda \in \mathbb{N}_0$ whereas \textit{fixed WMP} games require pseudo-polynomial time, i.e., also polynomial in the bound $\lambda$. Finally, multi-dimension \textit{bounded WMP} games are known to be non-primitive-recursive-hard and their decidability is still open~\cite{Chatterjee0RR15}. On the contrary, multi-dimension \textit{bounded WP} games are $\sf EXPTIME$-complete. This suggests that the colossal complexity of \textit{bounded WMP} games is a result of the quantitative nature of mean-payoff mixed with windows, and not an inherent drawback of the window mechanism.

\smallskip\noindent\textbf{Other related work.} In addition to the aforementioned articles, we mention two papers where logical formalisms dealing with time bounds are studied. In~\cite{DBLP:journals/fmsd/KupfermanPV09}, Kupferman et al.~introduced Prompt-LTL, which is strongly linked with the finitary conditions discussed above. In~\cite{DBLP:conf/csl/BaierKKW14}, Baier et al.~also studied an extension of LTL that can express properties based on the window mechanism of~\cite{Chatterjee0RR15}. The study of logical fragments corresponding to \textit{WP} games is an interesting question left open for future work.

\smallskip\noindent\textbf{Outline.} Section~\ref{sec:preliminaries} presents the needed definitions and known results about classical objectives. Section~\ref{sec:objectives} introduces the different objectives studied in this paper and establishes the links between them. Section~\ref{sec:onedim} and Section~\ref{sec:manydim} respectively present our results for one-dimension and multi-dimension games. Full proofs and detailed results, as well as additional discussion of related topics, can be found in the full version of this paper on arXiv~\cite{DBLP:journals/corr/BruyereHR16}.
\section{Preliminaries} \label{sec:preliminaries}

\smallskip\noindent\textbf{Game structures.}
We consider zero-sum turn-based games played by two players, $\playerOne$ and $\playerTwo$, on a finite directed graph. A \emph{game structure} is a tuple $G = (V_1,V_2,E)$ where $(i)$ $(V,E)$ is a finite directed graph, with $V = V_1 \cup V_2$ the set of vertices and $E \subseteq V \times V$ the set of edges such that for each $v \in V$, there exists $(v,v') \in E$ for some $v' \in V$ (no deadlock), $(ii)$ $(V_1,V_2)$ forms a partition of $V$ such that $V_i$ is the set of vertices controlled by player $\playerI$ with $i \in \{1,2\}$.

A \emph{play} of $G$ is an infinite sequence of vertices $\pi = v_0 v_1 \ldots \in V^{\omega}$ such that $(v_k,v_{k+1}) \in E$ for all $k \in \N$. We denote by $\Plays(G)$ the set of plays in $G$. \emph{Histories} of $G$ are finite sequences $\rho = v_0 \ldots v_k \in V^\ast$ defined in the same way. Given a play $\pi = v_0 v_1 \ldots$, the history $v_k \ldots v_{k+l}$ is denoted by $\pifactor{k}{k+l}$; in particular, $v_k = \pi[k]$. We also use notation $\pifactor{k}{\infty}$ for the suffix $v_k v_{k+1} \ldots$ of $\pi$.

\smallskip\noindent\textbf{Strategies.} A \emph{strategy} $\sigma_i$ for $\playerI$ is a function $\sigma_i\colon V^*V_i \rightarrow V$ assigning to each history $\rho v \in V^*V_i$ a vertex $v' = \sigma_i(\rho v)$ such that $(v,v') \in E$. It is \emph{finite-memory} if it can be encoded by a deterministic \emph{Moore machine}. The \textit{size} of the strategy is the size of its Moore machine. It is \emph{memoryless} if $\sigma_i(\rho v) = \sigma_i(\rho'v)$ for all histories $\rho v, \rho'v$ ending with the same vertex $v$, that is, if $\sigma_i$ is a function $\sigma_i\colon V_i \rightarrow V$.

Given a strategy $\sigma_i$ of $\playerI$, we say that a play $\pi = v_0 v_1 \ldots$ of $G$ is \emph{consistent} with $\sigma_i$ if $v_{k+1} = \sigma_i(v_0 \ldots v_k)$ for all $k \in \N$ such that $v_k \in V_i$. Consistency is naturally extended to histories in a similar fashion. Given an \emph{initial vertex} $v_0$, and a strategy $\sigma_i$ of each player $\playerI$, we have a unique play consistent with both strategies. This play is called the \emph{outcome} of the game and is denoted by $\Out(v_0,\sigma_1,\sigma_2)$.

\smallskip\noindent\textbf{Objectives and winning sets.} Let $G = (V_1,V_2,E)$ be a game structure. An \emph{objective for $\playerOne$} is a set of plays $\Obj \subseteq \Plays(G)$. A play $\pi$ is \emph{winning} for $\playerOne$ if $\pi \in \Obj$, and losing otherwise (i.e., winning for $\playerTwo$). We thus consider \emph{zero-sum} games such that the objective of player $\playerTwo$ is $\overline{\Obj} = \Plays(G) \ssetminus \Obj$. In the following, we always take the point of view of $\playerOne$ by assuming that $\Obj$ is his objective, and we denote by $(G,\Obj)$ the corresponding \emph{game}. Given an initial vertex $v_0$ of a game $(G,\Obj)$, a strategy $\sigma_1$ for $\playerOne$ is \emph{winning} from $v_0$ if $\Out(v_0,\sigma_1,\sigma_2) \in \Obj$ for all strategies $\sigma_2$ of $\playerTwo$. Vertex $v_0$ is also called \emph{winning} for $\playerOne$ and the \emph{winning set} $\WinG{1}{\Obj}{G}$ is the set of all his winning vertices. Similarly the winning vertices of $\playerTwo$ are those from which $\playerTwo$ can ensure to satisfy his objective $\overline{\Obj}$ against all strategies of $\playerOne$, and $\WinG{2}{\overline{\Obj}}{G}$ is his winning set. If $\WinG{1}{\Obj}{G} \cup \WinG{2}{\overline{\Obj}}{G} = V$, we say that the game is \emph{determined}. It is known that every turn-based game with a Borel objective is determined~\cite{Martin75}. This in particular applies to the objectives studied in this paper.

\smallskip\noindent\textbf{Decision problem.} Given a game $(G, \Obj)$ and an initial vertex $v_0$, we want to decide whether $\playerOne$ has a winning strategy from $v_0$ for the objective $\Obj$ or not (in which case, $\playerTwo$ has one for $\overline{\Obj}$). We want to study the complexity class of this decision problem as well as the memory requirements of winning strategies of both players. In this paper, we focus on several variants of the parity objective and we consider two settings: the \emph{one-dimension} case with one objective~$\Obj$ and the \emph{multi-dimension} case with the intersection of several objectives $\cap_{m=1}^n \Obj_m$.

\smallskip\noindent\textbf{Parity objective.} 
Let $G$ be a game structure. Let $\pi$ be a play, we define $\Inf(\pi)$ as the set of vertices seen infinitely often in $\pi$. Formally, $\Inf(\pi) = \{v \in V \mid \forall\, k \geq 0,\: \exists\, l \geq k,\: \pi[l] = v \}$.

Given a \emph{priority function} $p\colon V \rightarrow \{0,1,\ldots, d\}$ that maps every vertex to an integer priority where $d$ is even and $d \leq |V| +1$ (w.l.o.g.), the \emph{parity objective} $\Par(p)$ asks that of the vertices that are visited infinitely often, the smallest priority be even. Formally, the parity objective is defined as $\Par(p)= \{ \pi \in \Plays(G) \mid \min_{v \in \Inf(\pi)} p(v) \text{ is even} \}$. As smallest even priorities have a specific role in parity objectives, we define a partial order $\stronger$ on priorities as follows. For $c,c' \in \{0,\ldots,d\}$, we have $c \stronger c'$ if and only if $c$ is even and $c \leq c'$. In this case we say that $c$ is \smaller\ than $c'$. State-of-the-art results about parity games were already discussed in Section~\ref{sec:intro}.

\smallskip\noindent\textbf{Other useful objectives.} We recall some useful results for several classical objectives. Let $G$ be a game structure and $U \subseteq V$ be a set of vertices. A \emph{reachability} objective $\Reach(U)$ asks to visit a vertex of $U$ at least once, whereas a \emph{safety} objective $\Safe(U)$ asks to visit no vertex of $V \setminus U$. Deciding the winner in reachability games and safety games is known to be $\sf P$-complete with an algorithm in time $O(|V| + |E|)$, and memoryless winning strategies suffice for both objectives and both players~\cite{Beeri80,2001automata,Immerman81}. A \emph{B\"uchi} objective $\Buchi(U)$ asks to visit a vertex of $U$ infinitely often, whereas a \emph{co-B\"uchi} objective $\CoBuchi(U)$ asks to visit no vertex of $V \setminus U$ infinitely often. Deciding the winner in B\"uchi games and co-B\"uchi games is also $\sf P$-complete with an algorithm in time $O(|V|^2)$, and memoryless strategies also suffice for both objectives and both players~\cite{ChatterjeeH14,EmersonJ91,2001automata}.

\section{Adding time bounds to parity games} \label{sec:objectives}

In this section, we introduce the two approaches discussed in this paper: \textit{window parity} (\textsf{WP}) and \textit{parity-response} (\textsf{PR}) games.

\smallskip\noindent\textbf{Window parity and parity-response objectives.} The intuition for both approaches is as follows. The parity-response objective asks that every priority be followed by a \smaller\ priority in a bounded number of steps. In a window parity game, a \emph{window} with a bounded size is sliding along the play, and one asks to find a \smallest\footnote{Notice the difference: smallest vs.~smaller.} priority inside this window, and this for all positions along the play. We derive four variants for each of these objectives, according to whether the bound is given as a parameter or not (\emph{fixed} or \emph{bounded} variant), and whether the objective must be satisfied directly or eventually (\emph{direct} or \emph{undirect} variant). The \textit{undirect} variants are thus prefix-independent.\footnote{An objective $\Obj$ is \textit{prefix-independent} if for any play $\pi = \rho \pi'$, it holds that $\pi \in \Obj \iff \pi' \in \Obj$.} Formally:

\begin{definition}\label{def:obj}
Given a game structure $G = \Gdev$, a priority function $p\colon V \rightarrow \{0,1,\ldots, d\}$, and a \emph{bound} $\lambda \in \Nzero$, we define the eight following objectives:
\begin{eqnarray*}
\DirPR(\lambda,p) &=& \{ \pi \in \Plays(G) \mid \forall\, j \geq 0,\; \exists\, l < \lambda,\; p(\pi[j+l]) \stronger p(\pi[j]) \},\\
\DirFWP(\lambda,p) &=& \{ \pi \in \Plays(G) \mid \forall\, j \geq 0,\; \exists\, l < \lambda,\; \forall\, k \leq l,\; p(\pi[j+l]) \stronger p(\pi[j+k]) \},
\end{eqnarray*}
and given ${\sf X} \in \{\sf{PR,WP}\},$
\begin{eqnarray*}
{\sf FixX}(\lambda,p) &=& \{ \pi \in \Plays(G) \mid \exists\, i \geq 0,\; \pi[i,\infty] \in {\sf DirFixX}(\lambda,p)\}, \\
{\sf DirBndX}(p) &=& \{ \pi \in \Plays(G) \mid \exists\, \lambda \in \Nzero,\; \pi \in {\sf DirFixX}(\lambda,p) \},\\
{\sf BndX}(p) &=& \{ \pi \in \Plays(G) \mid \exists\, i \geq 0,\; \pifactor{i}{\infty} \in {\sf DirBndX}(p) \}.
\end{eqnarray*}
\end{definition}

Thus, in the direct fixed parity-response objective $\DirPR(\lambda,p)$, for all positions $j \geq 0$, the priority $p(\pi[j])$ must be followed by a \smaller\ priority $p(\pi[j+l])$ within at most $\lambda-1$ steps. The undirect fixed variant $\PR(\lambda,p)$ asks this objective to be satisfied eventually (i.e., for all positions $j \geq i$, for some~$i$). The direct bounded variant $\DirBP(p)$ (resp.~the undirect bounded variant $\BP(p)$) asks for the existence of a bound $\lambda$ for which $\DirPR(\lambda,p)$ (resp.~$\PR(\lambda,p)$) is satisfied.

For window parity objectives, we call $\lambda$ the \emph{window size}. Given a play $\pi = v_0v_1 \ldots$, a \emph{$\lambda$-window at position $j$} is a window of size $\lambda$ placed along $\pi$ from position $j$ to $j + \lambda-1$. The direct fixed window parity objective $\DirFWP(\lambda,p)$ asks that for all $j \geq 0$, inside the $\lambda$-\window{j}, one can find a priority $p(\pi[j+l])$ with $l \leq \lambda - 1$ that is the \smallest\ one in $\pifactor{j}{j+l}$. When the window size $\lambda$ is clear from the context, we drop the prefix $\lambda$ and simply talk about windows instead of $\lambda$-windows.

\begin{figure}[t]
\begin{minipage}[T]{.45\linewidth}
\centering
  \begin{tikzpicture}[scale=5]
    \everymath{\footnotesize}
    \draw (0,0) node [circle, draw] (A) {$v_0$};
    \draw (0.3,0) node [circle, draw] (B) {$v_1$};
    \draw (0.6,0) node [circle, draw] (C) {$v_2$};
    \draw (0.9,0) node [circle, draw] (G) {$v_3$};
    
    \draw (0,-.12) node[] (D) {$3$};
	\draw (0.3,-.12) node[] (E) {$1$};    
    \draw (0.6,-.12) node[] (F) {$2$};
    \draw (0.9,-.12) node[] (H) {$0$};
    
    \draw[->,>=latex] (A) to (B);
    \draw[->,>=latex] (B) to (C);
    \draw[->,>=latex] (C) to (G);
    
    \draw[->,>=latex] (G) to[bend right] (A);
    
	\path (-0.2,0) edge [->,>=latex] (A);    
    
    \end{tikzpicture}
    \vspace{-3mm}
    \caption{A simple example of one-player-game: all vertices belong to $\playerOne$.} 
\label{fig:Firstexample}
\end{minipage}\hfill
\begin{minipage}[T]{.45\linewidth}
\centering
  \begin{tikzpicture}[scale=5]
    \everymath{\footnotesize}
    \draw (0,0) node [circle, draw] (A) {$v_0$};
    \draw (0.45,0) node [rectangle, draw,inner sep=4pt] (B) {$v_1$};
    \draw (0.9,0) node [circle, draw] (C) {$v_2$};
    
    \draw (0,-.12) node[] (D) {$1$};
	\draw (0.45,-.12) node[] (E) {$2$};    
    \draw (0.9,-.12) node[] (F) {$0$};
    
    \draw[->,>=latex] (A) to (B);
    \draw[->,>=latex] (B) to (C);
    
    \draw[->,>=latex] (C) to[bend right] (A);
    \draw[->,>=latex] (B) .. controls +(315:0.3cm) and +(225:0.3cm) .. (B);
    
	\path (-0.2,0) edge [->,>=latex] (A);    
    
    \end{tikzpicture}
    \vspace{-5mm}
\caption{A game where $\playerOne$ wins for parity but loses for all variants of objectives \textsf{WP} and \textsf{PR}.}
\label{fig:Secondexample}
\end{minipage}
\end{figure}

\begin{example} \label{ex:1} We illustrate the definitions on a simple example where all vertices belong to $\playerOne$ (Figure~\ref{fig:Firstexample}). In this example and in the sequel, the priority $p(v)$ is always put under vertex~$v$, and circle (resp. square) vertices all belong to $\playerOne$ (resp. $\playerTwo$).
In the game of Figure~\ref{fig:Firstexample}, there is a unique play from the initial vertex $v_0$ equal to $\pi = (v_0v_1v_2v_3)^{\omega}$. On the one hand, we have that $\pi \in \DirPR(\lambda,p)$ for $\lambda = 3$. Indeed, the odd priority $3$ (resp. $1$) is followed by the even priority $2$ (resp. $0$) in exactly $\lambda - 1 = 2$ steps, whereas the even priority $2$ (resp. $0$) is ``followed" by itself in $0$ steps. Similarly, $\pi$ also belongs to the three variants $\PR(\lambda,p)$, $\DirBP(p)$ and $\BP(p)$. On the other hand, $\pi \not\in \DirFWP(3,p)$. Indeed, in the $3$-\window{0}, there is no $l \in \{0,1,2\}$ such that $p(v_l)$ is the \smallest\ priority in $\pifactor{0}{l}$ because $p(v_0)$ and $p(v_1)$ are odd, and $p(v_2)$ is even but $p(v_2) = 2 \not\stronger 1 = p(v_1)$. However, one can check that $\pi \in \DirFWP(4,p)$, and it also belongs to $\FWP(4,p)$, $\DirBWP(p)$ and $\BWP(p)$.\hfill$\triangleleft$
\end{example}

\smallskip\noindent\textbf{Relationship between objectives.} We now detail the inclusions and equalities between the various objectives introduced in Definition~\ref{def:obj} as well as with the parity objective.

\begin{proposition}
\label{prop:inclusions}
Let $G = (V_1, V_2, E)$ be a game structure and $p$ be a priority function. Let $\lambda \in \Nzero$.
\begin{enumerate}
\item For all ${\sf X} \in \{\PR(\lambda, p), \FWP(\lambda, p), \BP(p), \BWP(p)\}$, ${\sf DirX}$ $\subseteq {\sf X}$. \label{item:direct}
\item For all ${\sf X} \in \{\sf{PR},\sf{WP}\}$, $({\sf Dir}){\sf FixX}(\lambda, p)$ $\subseteq ({\sf Dir}){\sf BndX}(p)$. \label{item:bounded}
\item For all $\lambda' > \lambda$, for all ${\sf X} \in \{\PR,\FWP\}$, $({\sf Dir}){\sf X}(\lambda,p) \subseteq ({\sf Dir}){\sf X}(\lambda',p)$.\label{item:monotone}
\item ${\sf(Dir)}\FWP(\lambda,p) \subseteq {\sf(Dir)}\PR(\lambda,p)$. \label{item:WPcPR}
\item $({\sf Dir})\PR(\lambda,p) \subseteq ({\sf Dir})\FWP(\frac{d}{2} \cdot \lambda,p)$. \label{item:encadrement}
\item $({\sf Dir})\BP(p) = ({\sf Dir})\BWP(p)$.\label{item:samebnd}
\item ${\sf(Dir)}\BWP(p)\subseteq \Par(p)$. \label{item:BWPcPar}
\end{enumerate}
\end{proposition}

\begin{figure}[bht]
\centering
\begin{tikzpicture}[scale=3]

	\draw (-0.4,0) node (R) {$\ldots$};
	\draw (4.4,0) node (R) {$\ldots$};
	\draw (0,0) node [circle, draw,inner sep=1pt] (A) {};
	\draw (-0.05,0.08) node [] (AA) {$c_1$};
	\draw (1.05,0.08) node [] (AAb) {$c'_1$};
	\draw (0.25,0) node [circle, draw,inner sep=1pt] (B) {};
	\draw (0.5,0) node [circle, draw,inner sep=1pt] (C) {};
	\draw (0.75,0) node [circle, draw,inner sep=1pt] (D) {};
	\draw (0.75,-0.15) node [] (DD) {};
	\draw (1,0) node [circle, draw,inner sep=1pt] (E) {};
	\draw (1.25,0) node [circle, draw,inner sep=1pt] (F) {};
	\draw (1.5,0) node [circle, draw,inner sep=1pt] (G) {};
	\draw (1.75,0) node [circle, draw,inner sep=1pt] (H) {};
	\draw (2,0) node [circle, draw,inner sep=1pt] (I) {};
	\draw (2,-0.15) node [] (II) {};
	\draw (2.25,0) node [circle, draw,inner sep=1pt] (J) {};
	\draw (2.5,0) node [circle, draw,inner sep=1pt] (K) {};
	\draw (2.5,-0.15) node [] (KK) {};
	\draw (2.75,0) node [circle, draw,inner sep=1pt] (L) {};
	\draw (3,0) node [circle, draw,inner sep=1pt] (M) {};
	\draw (3.25,0) node [circle, draw,inner sep=1pt] (N) {};
	\draw (3.5,0) node [circle, draw,inner sep=1pt] (O) {};
	\draw (3.75,0) node [circle, draw,inner sep=1pt] (P) {};
	\draw (4,0) node [circle, draw,inner sep=1pt] (Q) {};
	\draw (4,-0.15) node [] (QQ) {};
	\draw (3,-0.2) node (R) {$\ldots$};
	\draw (1.45,0.08) node [] (l) {$c_3$};
	\draw (2.55,0.08) node [] (l) {$c'_3$};
	\draw (0.45,-0.08) node [] (l) {$c_2$};
	\draw (1.80,-0.08) node [] (l) {$c'_2$};
	\draw (2.2,-0.08) node [] (l) {$c_4$};
	\draw (3.80,-0.08) node [] (l) {$c'_k$};
	
	\draw[->,>=latex] (A) to[bend left=30] (E);
	\draw[->,>=latex,dashed] (C) to[bend right=35] (H);
	\draw[->,>=latex,dotted,thick] (G) to[bend left=30] (K);
	\draw (J) to[bend right=15] (2.8,-0.2);
	\draw[->,>=latex] (3.2,-0.2) to[bend right=15] (P);
	\draw (0.5,0.22) node (l1) {$\leq \lambda - 1$};
	\draw (2,0.22) node (l1) {$\leq \lambda - 1$};
	\draw (1.125,-0.3) node (l1) {$\leq \lambda - 1$};
	
	\draw (-0.3,0) -- (4.3,0);
	\draw (A) -- (B);
	\draw (B) -- (C);
	\draw (C) -- (D);
	\draw (D) -- (E);
	\draw (E) -- (F);
	\draw (F) -- (G);
	\draw (G) -- (H);
	\draw (H) -- (I);
	\draw (I) -- (J);
	\draw (J) -- (K);
	\draw (K) -- (L);
	\draw (L) -- (M);
	\draw (M) -- (N);
	\draw (N) -- (O);
	\draw (O) -- (P);
	\draw (P) -- (Q);

\end{tikzpicture}
\vspace{-3mm}
\caption{Illustration of inclusion $({\sf Dir})\PR(\lambda,p) \subseteq ({\sf Dir})\FWP(\frac{d}{2} \cdot \lambda,p)$.}
\label{fig:fromPRtoWP}
\end{figure}

We give here an intuitive explanation of Item~\ref{item:encadrement}, as it is the most interesting one technically. Assume we have a play $\pi \in \DirPR(\lambda,p)$, like the one depicted in Figure~\ref{fig:fromPRtoWP}. Since it satisfies objective $\DirPR(\lambda,p)$, we know that each odd priority is followed by a smaller even priority in at most $(\lambda-1)$ steps. We argue that it belongs to $\DirFWP(\frac{d}{2} \cdot \lambda,p)$, i.e., that for any position $j \geq 0$, the $(\frac{d}{2} \cdot \lambda)$-\window{j} sees a \smallest\ priority in some position $j+l$ with $l < \frac{d}{2} \cdot \lambda$. The key idea is depicted in Figure~\ref{fig:fromPRtoWP}.
Let $c_1$ be an odd priority. It must be followed by a \smaller\ priority $c'_1$ in at most $(\lambda - 1)$ steps. If $c'_1$ is the minimal priority encountered from $c_1$ to $c'_1$, then we are done. Assume it is not, then there exists $c_2$ between $c_1$ and $c'_1$ such that $c_2$ is odd and $c'_1 \not\stronger c_2$. But again, $c_2$ must be followed by $c'_2 \stronger c_2$ in at most $(\lambda - 1)$ steps. Repeating this argument, we obtain that $c_1$ is followed by a priority $c'_k$ in strictly less than $\frac{d}{2}\cdot \lambda$ steps\footnote{Actually, $\frac{d}{2}\cdot(\lambda-1) + 1$ but we use the simpler bound $\frac{d}{2}\cdot\lambda$ from now on for the sake of readability.} (as there are $\frac{d}{2}$ odd priorities and each of them is answered in $(\lambda-1)$ steps) such that $c'_k$ is even and smaller than all priorities encountered from $c_1$ to~$c'_k$. Therefore, $c'_k$ is the \smallest\ priority in $\pi[j, j+l]$ for some $l < \frac{d}{2}\cdot\lambda$. Since this argument can be repeated for any position $j \geq 0$, we obtain that the play satisfies $\DirFWP(\frac{d}{2}\cdot \lambda, p)$ as claimed.

From the inclusions $\Obj \subseteq \Obj'$ of Proposition~\ref{prop:inclusions}, we immediately derive the inclusions $\WinG{1}{\Obj}{G} \subseteq \WinG{1}{\Obj'}{G}$. It yields two interesting observations mentioned in Section~\ref{sec:intro}.
Notice that the inclusions of Proposition~\ref{prop:inclusions} are strict in general. This is also the case when one replaces the objectives by the winning sets of $\playerOne$ for these objectives. We briefly sketch the most interesting case here.

\begin{example}
\label{ex:shortEx}
Consider the game in Figure~\ref{fig:Secondexample}. The initial vertex $v_0$ is winning for the parity objective but is losing for all variants of objectives \textsf{WP} and \textsf{PR}: $\playerTwo$ has the possibility to use the self-loop on $v_1$ to delay for an arbitrarily long time the visit of the \smaller\ priority $0$ after seeing priority $1$, and can do so repeatedly using the other loop, thus defeating both direct and undirect variants of the objectives, as $\playerOne$ is never able to ensure a bound on the window size needed to see a \smallest\ priority. To win for the undirect bounded variants, $\playerTwo$ must use \textit{infinite memory} and play in rounds, increasing the time spent looping in $v_1$ at each round, thus preventing the existence of a bound.\hfill$\triangleleft$
\end{example}

We close this section by establishing that for the sub-case of games with priorities in $\{0, 1, 2\}$, \textsf{WP} and \textsf{PR} objectives coincide.

\begin{lemma}
\label{lem:sameobj}
Let $G$ be a game structure and $p \colon V \rightarrow \{0,1,2\}$ be a priority function. For all $\lambda \in \Nzero$, we have that $({\sf Dir})\PR(\lambda,p)  = ({\sf Dir})\FWP(\lambda,p)$.
\end{lemma}

\section{One-dimension games} \label{sec:onedim}

We begin our study of \textsf{WP} and \textsf{PR} objectives with one-dimension games: in this setting, there is a unique priority function $p$ and the objective $\Obj$ is a single objective $({\sf Dir}){\sf FixX}$ or $({\sf Dir}){\sf BndX}$ for ${\sf X} \in \{{\sf PR}, {\sf WP}\}$.

\smallskip\noindent\textbf{Bounded variants.} Recall that by Proposition~\ref{prop:inclusions}, the bounded variants are equivalent. Furthermore, it is already known that games with objective $({\sf Dir})\BP(p)$ are solvable in polynomial time~\cite{ChatterjeeHH09}. The next theorem sums up the complexity landscape for bounded variants and enrich it by proving $\sf P$-hardness for the associated decision problems. The result is obtained via a reduction from reachability games. In terms of memory requirements, $\playerOne$ can play without memory whereas Example~\ref{ex:shortEx} already illustrated that $\playerTwo$ requires infinite memory in general. The linear memory bound for $\playerTwo$ and the direct variant was established in~\cite{DBLP:journals/corr/abs-1207-0663}.

\begin{theorem}
\label{thm:finitaryparity} 
Let $G = (V_1,V_2,E)$ be a game structure, $v_0$ be an initial vertex, $p$ be a priority function, and $\Obj$ be the objective $\DirBP(p)$ or $\DirBWP(p)$ (resp. $\BP(p)$ or $\BWP(p)$). 
\begin{enumerate}
\item Deciding the winner in $(G,\Obj)$ from $v_0$ is $\sf P$-complete with an algorithm in $\mathcal{O}(|V| \cdot |E|)$ (respectively $\mathcal{O}(|V|^2 \cdot |E|)$) time, memoryless strategies are sufficient for $\playerOne$, and linear-memory strategies are necessary and sufficient for $\playerTwo$ (respectively infinite memory is necessary for $\playerTwo$). \label{item:finitaryparityItem1}
\item $\forall\,\lambda \geq |V|,\; \forall\, \lambda' \geq \frac{d}{2}\cdot|V|$, the winning sets for the objectives $\BP(p)$, $\PR(\lambda,p)$, $\BWP(p)$, and $\FWP(\lambda',p)$
are all equal. The same equalities hold for the direct variants ${\sf (Dir)}$.\label{item:finitaryparityItem2}
\end{enumerate}
\end{theorem}

The fixed variants are more interesting: the \textsf{PR} and \textsf{WP} approaches yield different results in this setting. We start with the \textsf{PR} one, for which we provide two polynomial-time algorithms for fixed-parameter sub-cases, hence significantly reducing the complexity of the problem (which is $\sf PSPACE$-complete in the general case).

\smallskip\noindent\textbf{Fixed parity-response objectives.}
Deciding the winner in $({\sf Dir})\PR$ games was very recently proved to be $\sf PSPACE$-complete~\cite{Weinert016}. As mentioned in Section~\ref{sec:intro}, the proof was actually provided for a more general model, but already holds for both $\PR$ and $\DirPR$ games.
Observe that the $\sf PSPACE$-hardness only holds for time bounds $\lambda < \vert V\vert$ since we know by Theorem~\ref{thm:finitaryparity}, Item~\ref{item:finitaryparityItem2}, that for larger values, the objectives are equivalent to the bounded variants, hence the corresponding decision problems lie in~$\sf P$.
We focus on the case $\lambda < |V|$: we show in the next theorem that when we fix either the largest priority~$d$ or the bound $\lambda$, the complexity collapses to~$\sf P$. We briefly sketch the corresponding algorithms here.

First, consider the case where $d$ is fixed. We reduce the $\PR(\lambda, p)$ (resp.~$\DirPR(\lambda,p)$) game to a co-B\"uchi (resp.~safety) game on an extended graph where we keep track of additional information in the vertices. Namely, we keep a vector that represents, for each odd priority $c$, the number of steps since seeing~$c$ without seeing any \smaller\ priority iboundsn the meantime. When this number reaches~$\lambda$ for any odd priority, we visit a special ``bad vertex'' and then reset the counters in the vector and resume the game. Essentially, winning for $\PR(\lambda, p)$ (resp.~$\DirPR(\lambda,p)$) boils down to eventually (resp.~completely) avoiding those bad vertices, hence to a co-B\"uchi (resp.~safety) game. This extended game has size $\mathcal{O}(|V|\cdot \lambda^{\frac{d}{2}})$ and can be solved in polynomial time since $\lambda < |V|$ and $d$ is fixed.

Second, consider the case where $\lambda$ is fixed. We also reduce the $\PR(\lambda, p)$ (resp.~$\DirPR(\lambda,p)$) game to a co-B\"uchi (resp.~safety) game, but with a different extended graph. Specifically, we here keep track of the last $\lambda$ vertices seen in the original game, and we want to avoid vertices of the extended graph that correspond to histories where an odd priority $c$ is not followed by a priority $c' \stronger c$ within $(\lambda-1)$ steps. Again, this can be expressed as either a co-B\"uchi or a safety objective depending on whether we are interested in the undirect or the direct variant respectively. The extended game has size $\mathcal{O}(|V|^\lambda)$ hence can be solved in polynomial time since $\lambda$ is fixed.

\begin{theorem}
\label{thm:fixed}
Let $G = (V_1,V_2,E)$ be a game structure, $v_0$ be an initial vertex, $p \colon V \rightarrow \{0,\ldots,d\}$ be a priority function, and $\Obj$ be the objective $\DirPR(\lambda,p)$ (resp. $\PR(\lambda,p)$) for some $\lambda < |V|$. If either $d$ is fixed or $\lambda$ is fixed, deciding the winner in $(G,\Obj)$ from $v_0$ is in $\sf P$. More precisely, if $d$ is fixed, deciding the winner can be done in $\mathcal{O}((|V| + |E|) \cdot \lambda^\frac{d}{2})$ (resp. $\mathcal{O}(|V|^2 \cdot \lambda^d)$) time, and if $\lambda$ is fixed, deciding the winner can be done in $\mathcal{O}((|V|+|E|)\cdot |V|^{\lambda-1})$ (resp. $\mathcal{O}(|V|^{2\lambda})$). In both cases, polynomial-memory strategies are sufficient for both players, and memory is necessary even in one-player games.
\end{theorem}

\smallskip\noindent\textbf{Fixed window parity objectives.}
Whereas $({\sf Dir})\PR$ games are {\sf PSPACE}-complete, we now establish that $({\sf Dir})\FWP$ games are {\sf P}-complete. 
Observe that if $\lambda \geq \frac{d}{2} \cdot |V|$, the problem boils down to solving the bounded variant thanks to Theorem~\ref{thm:finitaryparity}. Hence, we focus on the case where $\lambda < \frac{d}{2} \cdot |V|$. 

Our algorithm is inspired by the approach developed for \textit{window mean-payoff games} in~\cite{Chatterjee0RR15}. It can be sketched as follows. As for the fixed-parameter algorithms for $({\sf Dir})\PR$ games presented in Theorem~\ref{thm:fixed}, we want to reduce the $\FWP$ and $\DirFWP$ games to co-B\"uchi and safety games respectively, where $\playerOne$ wants to avoid ``bad vertices'' representing a violation of the condition at stake. Here, such a violation represents a $\lambda$-\textit{bad window}, i.e., a window for which no even minimum priority is found before $\lambda$ steps. Detecting such $\lambda$-bad windows can be achieved by considering an extended game structure where we encode additional information for the minimum priority of the current window and the number of steps in this window. A ``bad vertex'' is visited whenever we reach the end of a $\lambda$-window with an odd minimum priority. If an even minimum is found, it is also a minimum for the windows at intermediate positions and the step counter is reset. The extended game has size $\mathcal{O}(|V|\cdot d \cdot \lambda)$, hence polynomial size since $\lambda < \frac{d}{2} \cdot |V|$. Therefore, we can solve it in polynomial time. This is in contrast to window mean-payoff games where the fixed variant requires \textit{pseudo}-polynomial time in general~\cite{Chatterjee0RR15}.

Upper bounds on the memory are obtained by construction of our reduction and we prove polynomial lower bounds in the extended version of this paper~\cite{DBLP:journals/corr/BruyereHR16}. 

\begin{theorem}
\label{prop:WPonedim}
Let $G = (V_1,V_2,E)$ be a game structure, $v_0$ be an initial vertex, $p$ be a priority function, and $\Obj$ be the objective $\DirFWP(\lambda,p)$ (resp. $\FWP(\lambda,p)$) for some $\lambda < |V|$. Then deciding the winner in $(G,\Obj)$ from $v_0$ is {\sf P}-complete with an algorithm in $\mathcal{O}((|V| + |E|) \cdot d \cdot \lambda)$ (resp. $\mathcal{O}(|V|^2 \cdot d^2 \cdot \lambda^2)$) time. Polynomial-memory strategies are both sufficient and necessary for both players.
\end{theorem}

\section{Multi-dimension games} \label{sec:manydim}

We now consider multi-dimension games: in this setting, there are $n$ priority functions $p_1$, \ldots{}, $p_n$ and the objective $\Obj$ is the \textit{conjunction} of \textit{identical} objectives $\Obj_m$ for each ``dimension'' (i.e., priority function).

\subsection{Bounded variants} \label{subsec:multibounded}

Recall that Proposition~\ref{prop:inclusions} established the equality of objectives $({\sf Dir})\BWP(p)$ and $({\sf Dir})\BP(p)$ in the one-dimension setting. This equality trivially carries over to the multi-dimension setting, i.e., we have that $ \cap_{m=1}^n ({\sf Dir})\BWP(p_m) = \cap_{m=1}^n ({\sf Dir})\BP(p_m)$ since the individual objectives (one per priority function) are equal. Hence, it suffices to obtain our results for either \textsf{WP} or \textsf{PR} objectives.

\smallskip\noindent\textbf{Overview.} The next theorem presents an overview of our results. For a comparison of those results with related models, see Section~\ref{sec:intro}. We sketch the key points to prove the theorem in the following paragraphs.

\begin{theorem}
\label{prop:multiDirBP} 
Let $G = (V_1,V_2,E)$ be a game structure, $v_0$ be an initial vertex, $p_1,\ldots,p_n$ be $n$ priority functions, and $\Obj$ be the objective $\cap_{m=1}^n \DirBP(p_m)$ or $\cap_{m=1}^n \DirBWP(p_m)$ (resp. $\cap_{m=1}^n \BP(p_m)$ or $\cap_{m=1}^n \BWP(p_m)$). Let $b = |V| \cdot 2^{n\cdot\frac{d}{2}} \cdot n\cdot\frac{d}{2}$.
\begin{enumerate}
\item Deciding the winner in $(G,\Obj)$ from $v_0$ is $\sf EXPTIME$-complete with an algorithm in $\mathcal{O}(b^2)$ (resp. $\mathcal{O}(|V| \cdot b^2)$) time, and exponential-memory strategies are necessary and sufficient for both players (resp. for $\playerOne$ and infinite-memory is necessary for $\playerTwo$).
\vspace{1mm}
\item $\forall\, \lambda \geq b$, $\forall\, \lambda' \geq b\cdot\frac{d}{2}$, the winning sets for the following objectives are all equal: $\cap_{m=1}^n \BP(p_m)$, $\cap_{m=1}^n \PR(\lambda,p_m)$, $\cap_{m=1}^n \BWP(p_m)$, and $\cap_{m=1}^n \FWP(\lambda',p_m)$. The same equalities hold for the direct variants ${\sf (Dir)}$.\label{item:multidimequalities}
\end{enumerate}
\end{theorem}

\smallskip\noindent\textbf{Exponential-time algorithm and upper bounds on memory.} To prove ${\sf EXPTIME}$-membership, we introduce related games from the literature. First, \textit{request-response games}~\cite{WallmeierHT03,ChatterjeeHH09}. Consider $r$ sets of vertices $Rq_1,\ldots, Rq_r$ representing requests and $r$ sets of vertices $Rp_1,\ldots,Rp_r$ representing corresponding responses ($Rq_i, Rp_i \subseteq V$ for all $i$). 
The \emph{request-response} objective, denoted by $\RR((Rq_i,Rp_i)_{i=1}^r)$, requires that \textit{for all} $i$, whenever a vertex of $Rq_i$ is visited, then, later on, a vertex of $Rp_i$ is also visited.\footnote{Note that a single response $Rp_i$ suffices to answer all pending requests $Rq_i$, in the same spirit as for priorities in the \textit{parity-response} objective.} Observe that by definition, this objective is \textit{direct}, i.e., the condition must hold from the start, not only eventually. Several variants of these games have been studied in the literature, under various names. Those variants include \emph{direct bounded request-response} games: the objective $\DirBndRR((Rq_i,Rp_i)_{i=1}^r)$ asks that there exist a bound $b \in \mathbb{N}_0$ such that if a request is visited, then the corresponding response is visited within $b$ steps. Its undirect variant $\BndRR$ have also been considered. The results of interest for us are $(i)$ the $\sf EXPTIME$-membership of all those variants, $(ii)$ that exponential memory suffices for both players in all variants except for $\playerTwo$ and $\BndRR$ where infinite memory is necessary, and $(iii)$ that whatever the game~$G$, if $\playerOne$ can win for any of those objectives, he can ensure that eventually all requests are answered in at most $b = |V| \cdot 2^r \cdot r$ steps~\cite{WallmeierHT03,ChatterjeeHH09}.

We establish a polynomial-time reduction from multi-dimension $\DirBWP$ and $\BWP$ games (or equivalently, $\DirBP$ and $\BP$ games) to $\DirBndRR$ and $\BndRR$ games respectively. The crux is to model each odd priority in each dimension as a request whose corresponding response is the occurrence of a \smaller\ priority in the same dimension. Since we have $\frac{d}{2}$ odd priorities and $n$ dimensions, we need $n\cdot\frac{d}{2}$ pairs of requests and responses. We thus obtain an exponential-time algorithm for multi-dimension ${\sf (Dir)}\BWP$ and ${\sf (Dir)}\BP$ games, along with exponential upper bounds on memory in all cases except the one of $\playerTwo$ in $\BP$ and $\BWP$ games.

\smallskip\noindent\textbf{Equalities between objectives.} The key ingredient for the last item of  Theorem~\ref{prop:multiDirBP} is the aforementioned bound given in~\cite{ChatterjeeHH09} for $({\sf Dir})\BndRR$ games, and by extension, for $({\sf Dir})\BP$ and $({\sf Dir})\BWP$ games thanks to our reduction. The rest follows the same lines as in the one-dimension case, i.e., it builds upon the inclusions and equalities presented in Proposition~\ref{prop:inclusions}.

\smallskip\noindent\textbf{Lower bound on complexity.} To prove the $\sf EXPTIME$-hardness of objective $({\sf Dir})\BWP$ (and equivalently, of objective $({\sf Dir})\BP$), we establish a reduction from the \textit{membership problem for alternating polynomial-space Turing machines (APTMs)}~\cite{DBLP:journals/jacm/ChandraKS81}. Our proof is adapted from the reduction presented in~\cite[Lemma 23]{Chatterjee0RR15} in the related context of window mean-payoff games. Since technical details are similar to~\cite[Lemma 23]{Chatterjee0RR15}, we only include a high-level sketch of the reduction in the full version of this paper~\cite{DBLP:journals/corr/BruyereHR16}. The main change is the way we deal with windows: whereas weights were used for window mean-payoff games, we need here to emulate the same actions with adapted priorities.
Interestingly, our proof also shows $\sf EXPTIME$-hardness of the fixed variants, $({\sf Dir})\FWP$ and $({\sf Dir})\PR$. Furthermore, the hardness already holds with only three priorities ($d = 2$).

\smallskip\noindent\textbf{Lower bounds on memory.} The last missing pieces to the proof of Theorem~\ref{prop:multiDirBP} are the exponential lower bounds on memory. Recall that for $\playerTwo$ in \textit{undirect} bounded \textsf{WP} or \textsf{PR} games, we already proved that infinite memory is necessary in Example~\ref{ex:shortEx}. To cover all remaining cases and establish exponential lower bounds matching the upper bounds obtained above, we prove a polynomial-time reduction from \textit{generalized reachability games}~\cite{FijalkowH13} to multi-dimension $({\sf Dir})\BWP$ and $({\sf Dir})\BP$ games. Given $U_1,\ldots,U_n$ a family of $n$ subsets of $V$, a generalized reachability objective $\GenReach(U_1, . . . , U_n) =  \cap_{m=1}^n \Reach(U_m)$ asks to visit a vertex of $U_m$ at least once, for each $m \in \{1,\ldots,n\}$.
Since $\GenReach$ games are known to require exponential memory for both players~\cite{FijalkowH13}, the reduction yields the desired lower bounds. A similar reduction is presented for window mean-payoff games in~\cite{Chatterjee0RR15}. Interestingly, the same technique also works for multi-dimension $({\sf Dir})\FWP$ and $({\sf Dir})\PR$ games.

Let us sketch the reduction from $\GenReach$ to multi-dimension $\FWP$ games (the other cases are similar). Intuitively, if the generalized reachability objective asks to visits $n$ different target sets, we will use $n$ dimensions. We create a modified version of the game structure such that, at the start of the game, we see priority $1$ in all dimensions, and such that the only way to see priority $0$ in dimension $m \in \{1, \ldots{}, n\}$ is to visit the $m$-th target set. We also modify the game by giving $\playerTwo$ the possibility to force seeing $0$ in all dimensions and restart the game by seeing $1$'s again: this is necessary to ensure that the prefix-independence of objective $\FWP$ cannot help $\playerOne$ to win without visiting all target sets. Finally, we use the fact that if $\playerOne$ has a winning strategy in a $\GenReach$ game with $n$ targets, then he has one that wins in strictly less than $n \cdot |V|$ steps (i.e., edges), to define an appropriate window size $\lambda = 2\cdot n \cdot |V|$ for which the reduction to objective $\cap_{m=1}^n \FWP(\lambda,p_m)$ on our modified game structure holds. As in the reduction for $\sf EXPTIME$-hardness, we only need three priorities here ($d=2$). 

For the reader's interest, we complement this reduction with an example illustrating the need for exponential memory for $\playerOne$ in $\FWP$ games (it also works for the other objectives).

  \vspace{2mm}
\begin{figure}[htb]
  \centering   
  \scalebox{0.8}{\begin{tikzpicture}[->,>=stealth',shorten >=1pt,auto,node
    distance=2.5cm,bend angle=45,scale=0.4, font=\small]
    \tikzstyle{p1}=[draw,circle,text centered,minimum size=8mm]
    \tikzstyle{p2}=[draw,rectangle,text centered,minimum size=8mm]
    \node[p2]  (0)  at (0, 0) {$v_{1}$};
    \node[p2]  (1) at (4, 2) {$v_{1,L}$};
    \node[p2]  (2) at (4, -2)  {$v_{1,R}$};
    \node[p2]  (3) at (8, 0)  {$v_{n}$};
    \node[p2]  (4)  at (12, 2) {$v_{n,L}$};
    \node[p2]  (5)  at (12, -2) {$v_{n,R}$};
    \node[p1]  (6)  at (16, 0) {$u_{1}$};
    \node[p1]  (7) at (20, 2) {$u_{1,L}$};
    \node[p1]  (8) at (20, -2)  {$u_{1,R}$};
    \node[p1]  (9) at (24, 0)  {$u_{n}$};
    \node[p1]  (10)  at (28, 2) {$u_{n,L}$};
    \node[p1]  (11)  at (28, -2) {$u_{n,R}$};
    \coordinate[shift={(-5mm,0mm)}] (init) at (0.west);
    \path
    (init) edge (0);
    \draw[->,>=latex] (0) to (1);
    \draw[->,>=latex] (0) to (2);
    \draw[->,>=latex] (3) to (4);
    \draw[->,>=latex] (3) to (5);
    \draw[dotted,->,>=latex] (1) to (3);
    \draw[dotted,->,>=latex] (2) to (3);
    \draw[->,>=latex] (4) to (6);
    \draw[->,>=latex] (5) to (6);
    \draw[->,>=latex] (6) to (7);
    \draw[->,>=latex] (6) to (8);
    \draw[dotted,->,>=latex] (7) to (9);
    \draw[dotted,->,>=latex] (8) to (9);
    \draw[->,>=latex] (9) to (10);
    \draw[->,>=latex] (9) to (11);
    
    \draw[->,>=latex] (10) to (28,4) to (0,4) to (0);
    \draw[->,>=latex] (11) to (28,-4) to (0,-4) to (0);
      \end{tikzpicture}}
      \caption{Family of multi-dimension games requiring exponential memory for $\playerOne$ for objective $\FWP$ with $\lambda = 3n$.}
      \label{fig:multiDimExpFamily}
  \end{figure}
  
\begin{example}
Consider the family of game structures depicted in Figure~\ref{fig:multiDimExpFamily}. This family is parameterized by $n \in \mathbb{N}_0$ and is inspired by a similar one proposed in~\cite{CRR14} for a different context (i.e., energy games). For each of these games structures, the number of vertices is linear in~$n$ ($|V| = 6n$) and we define $2n$ priority functions in the following way: for all $i \in \{1, \ldots, n\}$ and for all $m \in \{1, \ldots,  2n\}$, $p_m(v_i) = p_m(u_i) = 2$, 
\begin{eqnarray*}
 &p_m(v_{i,L}) = 
\begin{cases}
1 & \mbox{if $m = 2i-1$} \\
2 & \mbox{otherwise}
\end{cases},\ & \qquad p_m(v_{i,R}) =
\begin{cases}
1 & \mbox{if $m = 2i$} \\
2 & \mbox{otherwise}
\end{cases},\\
& p_m(u_{i,L}) = 
\begin{cases}
0 & \mbox{if $m = 2i-1$} \\
2 & \mbox{otherwise}
\end{cases},\ & \qquad p_m(u_{i,R}) = 
\begin{cases}
0 & \mbox{if $m = 2i$} \\
2 & \mbox{otherwise}
\end{cases}.\\
\end{eqnarray*}
Let $\Obj = \cap_{m=1}^{2n} \FWP(3n, p_m)$ be the objective of $\playerOne$. In order to prevent $(3n)$-bad windows, $\playerOne$ has to choose $u_{i,L}$ (resp.~$u_{i,R}$) whenever $\playerTwo$ chooses $v_{i,L}$ (resp.~$v_{i,R}$). Hence in order to prevent outcomes with infinitely-many $(3n)$-bad windows, $\playerOne$ must be able to record $2^n$ different histories from $v_1$ to $u_1$. This obviously requires exponential memory in $n$, hence in the size of the game.\hfill$\triangleleft$

\end{example}

\subsection{Fixed variants} \label{subsec:multifixed}

As in one-dimension, some differences arise for the fixed variants. See Section~\ref{sec:intro} for a comparison.

\smallskip\noindent\textbf{Parity-response objectives.} To establish an exponential-time algorithm for multi-dimension $\DirPR$ (resp.~$\PR$) games, we reduce them to safety (resp.~co-B\"uchi) games on an exponentially-larger game structure. Our reduction is in the same spirit\footnote{Note that the other algorithm suggested in Theorem~\ref{thm:fixed} and exponential in $\lambda$ is not interesting here, since $\lambda$ can be exponential before the fixed variant becomes equivalent to the bounded one (Theorem~\ref{prop:multiDirBP}), hence this algorithm would take \textit{doubly}-exponential time.} as the one for Theorem~\ref{thm:fixed} for the case $d$ fixed in one-dimension. That is, the extended structure encodes for each odd priority in each dimension, the number of steps since seeing the odd priority without seeing a \smaller\ priority in the meantime. The complexity and memory lower bounds follow from the reductions sketched for the bounded variants.

\begin{theorem}
\label{prop:multiPR} 
Let $G = (V_1,V_2,E)$ be a game structure, $v_0$ be an initial vertex, $p_1,\ldots,p_n$ be $n$ priority functions, and $\Obj$ be the objective $\cap_{m=1}^n \DirPR(\lambda,p_m)$ (resp. $\cap_{m=1}^n \PR(\lambda,p_m)$) for $\lambda \in \Nzero$. Deciding the winner in $(G,\Obj)$ from $v_0$ is $\sf EXPTIME$-complete with an algorithm in $\mathcal{O}((|V|+|E|) \cdot \lambda^{\frac{d}{2}\cdot n})$ (resp. $\mathcal{O}(|V|^2 \cdot \lambda^{d\cdot n})$) time. Exponential memory is both sufficient and necessary for both players.
\end{theorem}

\smallskip\noindent\textbf{Window parity objectives.} To conclude our study of \textsf{PR} and \textsf{WP} objectives, it remains to establish an exponential-time algorithm for multi-dimension $\DirFWP$ (resp.~$\FWP$) games. Again, we reduce those games to safety (resp.~co-B\"uchi) games on an exponentially-larger game structure. Our reduction is here based on the one used in the one-dimension setting (Theorem~\ref{prop:WPonedim}). That is, the extended structure encodes, for each dimension, the minimum priority of the current window, and the number of steps in that window. The complexity and memory lower bounds follow from the reductions sketched for the bounded variants.

\begin{theorem}
\label{prop:multiWP}
Let $G = (V_1,V_2,E)$ be a game structure, $v_0$ be an initial vertex, $p_1,\ldots,p_n$ be $n$ priority functions, and $\Obj$ be the objective $\cap_{m=1}^n \DirFWP(\lambda,p_m)$ (resp. $\cap_{m=1}^n \FWP(\lambda,p_m)$) for $\lambda \in \Nzero$. Deciding the winner in $(G,\Obj)$ from $v_0$ is $\sf EXPTIME$-complete with an algorithm in $\mathcal{O}((|V|+|E|)\cdot (d \cdot \lambda)^n)$ (resp. $\mathcal{O}(|V|^2 \cdot (d \cdot \lambda)^{2\cdot n})$) time. Exponential memory is both sufficient and necessary for both players.
\end{theorem}

\paragraph*{\bf{Acknowledgments.}} We express our gratitude to Jean-Fran\c{c}ois Raskin (Universit\'e libre de Bruxelles, Belgium) and Martin Zimmermann (Saarland University, Germany) for insightful discussions.

\bibliographystyle{eptcs}
\bibliography{biblio}

\end{document}